\documentclass[10pt,preprint2]{aastex}
\usepackage{color}
\usepackage{url}
\usepackage{bm}

\shorttitle{Precursor Activities of the X1.6 flare in AR 12192}
\shortauthors{Bamba et al.}

\begin{document}

\title{Study on Precursor Activity of the X1.6 Flare\\in the Great AR 12192 with SDO, IRIS, and Hinode}
%\title{Study on the Relationship between the Plasma Flow and the Triggering Process of the X1.6 Flare in the great AR 12192}

\author{Yumi Bamba}
\affil{Institute of Space and Astronautical Science (ISAS)/Japan Aerospace Exploration Agency (JAXA)\\ 3-1-1 Yoshinodai, Chuo-ku, Sagamihara, Kanagawa 252-5210, Japan}
\email{y-bamba@nagoya-u.jp}

\author{Kyoung-Sun Lee}
\affil{National Astronomical Observatory of Japan (NAOJ)\\ 2-21-1 Osawa, Mitaka, Tokyo 181-8588, Japan}

\author{Shinsuke Imada}
\affil{Institute for Space-Earth Environmental Research (ISEE)/Nagoya University\\ Furo-cho, Chikusa-ku, Nagoya, Aichi 464-8601, Japan}

\author{Kanya Kusano}
\affil{Institute for Space-Earth Environmental Research (ISEE)/Nagoya University\\ Furo-cho, Chikusa-ku, Nagoya, Aichi 464-8601, Japan}

%%%ABSTRACT%%%
\begin{abstract} %less than 250 words

The physical properties and its contribution to the onset of solar flare are still unclear although chromospheric brightening is considered a precursor phenomenon of flare. Many studies suggested that photospheric magnetic field changes cause destabilization of large-scale coronal structure. We aim to understand how a small photospheric change contributes to a flare and to reveal how the intermediary chromosphere behaves in the precursor phase. We analyzed the precursor brightening of the X1.6 flare on 2014 October 22 in the AR 12192 using the Interface Region Imaging Spectrograph (IRIS) and Hinode/EUV Imaging Spectrometer (EIS) data. We investigated a localized jet with the strong precursor brightening, and compared the intensity, Doppler velocity, and line width in \ion{C}{2}, \ion{Mg}{2} k, \ion{Si}{4} lines by IRIS and \ion{He}{2}, \ion{Fe}{12}, \ion{Fe}{15} lines by Hinode/EIS. We also analyzed photospheric magnetic field and chromospheric/coronal structures using Solar Dynamics Observatory (SDO)/Helioseismic and Magnetic Imager (HMI) and Atmospheric Imaging Assembly (AIA). We found a significant blueshift ($\sim$100 $km~s^{-1}$), which is related to the strong precursor brightening over a characteristic magnetic field structure, and the blueshift was observed at all the temperature. This might indicate that the flow is accelerated by Lorentz force. Moreover, the large-scale coronal loop that connects the foot-points of the flare ribbons was destabilized just after the precursor brightening with the blueshift. It suggests that magnetic reconnection locally occurred in the lower chromosphere and it triggered magnetic reconnection of the X1.6 flare in the corona.

\end{abstract}


%%%INTRODUCTION%%%
\section{Introduction} \label{sec:intro}

Solar flares \citep{carrington59} are the biggest explosive phenomena in the solar system that releases magnetic energy stored in the solar corona mainly as thermal and kinetic energy of the coronal plasma.
Flares are observed as sudden enhancement of emission in multiple wavelength \citep{fletcher11}.
The CSHKP model \citep{carmichael64, sturrock66, hirayama74, kopppneuman76} proposed that magnetic reconnection drives magnetic energy release in a flare.
The model is widely accepted as the standard flare model, and it can explain observable features of flare such as a cusp-shaped structure in the corona \citep{tsuneta92}, a hard X-ray source above the flare loop-top \citep{masuda94}, plasmoid ejection \citep{ohyamashibata98}, reconnection inflow \citep{yokoyama01, chen04} and outflow \citep{imada13}.
However, there are remaining problems related to such as energy storage process, triggering process, elementary process of magnetic reconnection, and particle acceleration process.
In this study, we focus on the triggering process of flares.

Various models have been proposed for the triggering process of flares so far.
The kink mode destabilization model \citep{fangibson03, torokkliem05} in which the highly twisted flux rope erupts by helical kink instability have been proposed as a theoretical flare onset model.
The break-out model \citep{antiochos99}, in which the flux rope grows by magnetic reconnection and it ejects from overlying magnetic field by loss-of-equilibrium or loss-of-stability, is also known as a flare onset model.
According to these models, it is inferred that strong twist motion such as a photospheric horizontal flow to form field aligned electric current is observed.
An emerging magnetic flux \citep{chenshibata00} and tether-cutting magnetic reconnection \citep{moore01} are also proposed as an important precursor activity.
Moreover, a large number of studies have been investigated many kinds of precursory activities precede to a flare such as evolution of the coronal magnetic shear relating sunspot rotation \citep{su07, vemareddy12}, converging photospheric flow lead tether-cutting flux cancellation \citep{sterling10}, filament destabilization \citep{zuccarello09}, and coronal non-thermal flow relating photospheric helicity injection \citep{harra09}.
Therefore, over the last few decades, considerable effort has been devoted toward understanding the triggering process of flares and various flare models have been proposed.
However, it is still unclear what kind of ``trigger(s)'' is dominant in the actual solar surface although these models can explain a part of the physical processes of flares.
Moreover, observational verification of these models is insufficient, and the model that can universally explain solar flares is still not established.

Precursor brightening seen in multiple wavelength is widely considered as an important phenomenon prior to a flare.
\citet{tappin91} statistically studied the relation between flare occurrences and precursor brightening in soft and hard X-rays, and they found that the vast majority of X-ray flares are preceded by soft X-ray precursors some 10 to 60 minutes before the flare.
On the other hand, \citet{farnik98} pointed out that the physical properties of the flare sites found no consistent feature distinguishable from other non-flaring active region emission and hence no definite evidence of a special precursor phase in flares.
There are actually many brightening which is not directly related to flare triggering mechanism as they suggested.
In order to identify the precursor brightening which can be an indicator of a flare, it is mandatory to simultaneously analyze photospheric magnetic field data.
\citet{chifor07} and \citet{joshi11} analyzed flares corresponding to filament eruptions using hard X-ray, UV/EUV images, and magnetic field data.
They found that X-ray precursors provide evidence for localized magnetic reconnection which plays a crucial role in destabilizing the active region filaments leading to the flares.
\citet{bamba13} have studied precursor brightening seen from two hours to five minutes prior to the flare onset in chromospheric images.
They found that chromospheric precursor brightening is related to characteristic magnetic field structures that plays an important role for flare triggering.
These magnetic field structures were quantitatively consistent with theoretical prediction by \citet{kusano12}.
%These studies suggest that a {\bf precursor} brightening in the solar atmosphere can be an indicator of a flare.
However, the physical properties and its contribution of precursor brightening to large-scale energy release in a flare are still unclear.
Therefore, it is required to analyze the plasma dynamics in the phase between precursor brightening and flare onset in detail.
The spectroscopic observations in UV/EUV are useful to diagnose the plasma property from chromosphere to corona. 
Although there are several studies to diagnose the plasma during the precursor to pre-flare phase by spectroscopic observation \citep[e.g.,][]{brosiusphilips04, imada14}, most of them do not focus on the plasma condition at the sight of precursor brightening.
%It is also important to understand how a small-scale change of the magnetic field in the photosphere affects large-scale coronal structure.
%Further, understanding the behavior of plasma in the chromosphere, which is located between photosphere and corona, during the {\bf precursor} phase is also important.

In this study, we aim to clarify the physical properties and significance of precursor brightening in a onset process of a sample event.
We analyzed the X1.6 flare occurred in the great active region (AR) NOAA 12192 on 2014 October 22.
The AR was the biggest sunspot region in the solar cycle 24, and its magnetic structure was a $\delta$-type sunspot in which several umbra are sharing penumbra.
Generally $\delta$-type region has high possibility of a flare, and actually more than 130 flares, these are larger than C-class and including six X-class flares, occurred during its disk passage.
The X1.6 flare was the second X-class flare in the AR, and whole of the X1.6 flare was observed by both Hinode \citep{kosugi07}, Solar Dynamics Observatory (SDO; \citet{pesnell12}), Interface Region Imaging Spectrograph (IRIS; \citet{depontieu14}) and Reuven Ramaty High-Energy Solar Spectroscopic Imager (RHESSI; \citet{lin02}).
Especially, the central region of the flare ribbons was well covered by IRIS and Hinode/EUV Imaging Spectrometer (EIS; \citet{Culhane07}).
There are also magnetic field data and imaging data of the solar atmosphere by SDO.
We investigate the temporal evolution of the photospheric magnetic field structures and the chromospheric/coronal responses in the precursor phase of the X1.6 event, and discuss precursor activity and onset scenario of the flare.

This paper is organized as follows.
The data and analysis method are described in Sections 2 and 3, respectively.
The results from SDO, Hinode, and IRIS analysis are shown in Section 4.
The interpretation of the results and conceivable flare onset scenario of the X1.6 event are discussed in Section 5.
Finally, we summarize the result and significance of the study in Section 6.

%%%DATA DESCRIPTION%%%
\section{Data Description} \label{sec:data}

\begin{figure*}
\epsscale{1.80}
\plotone{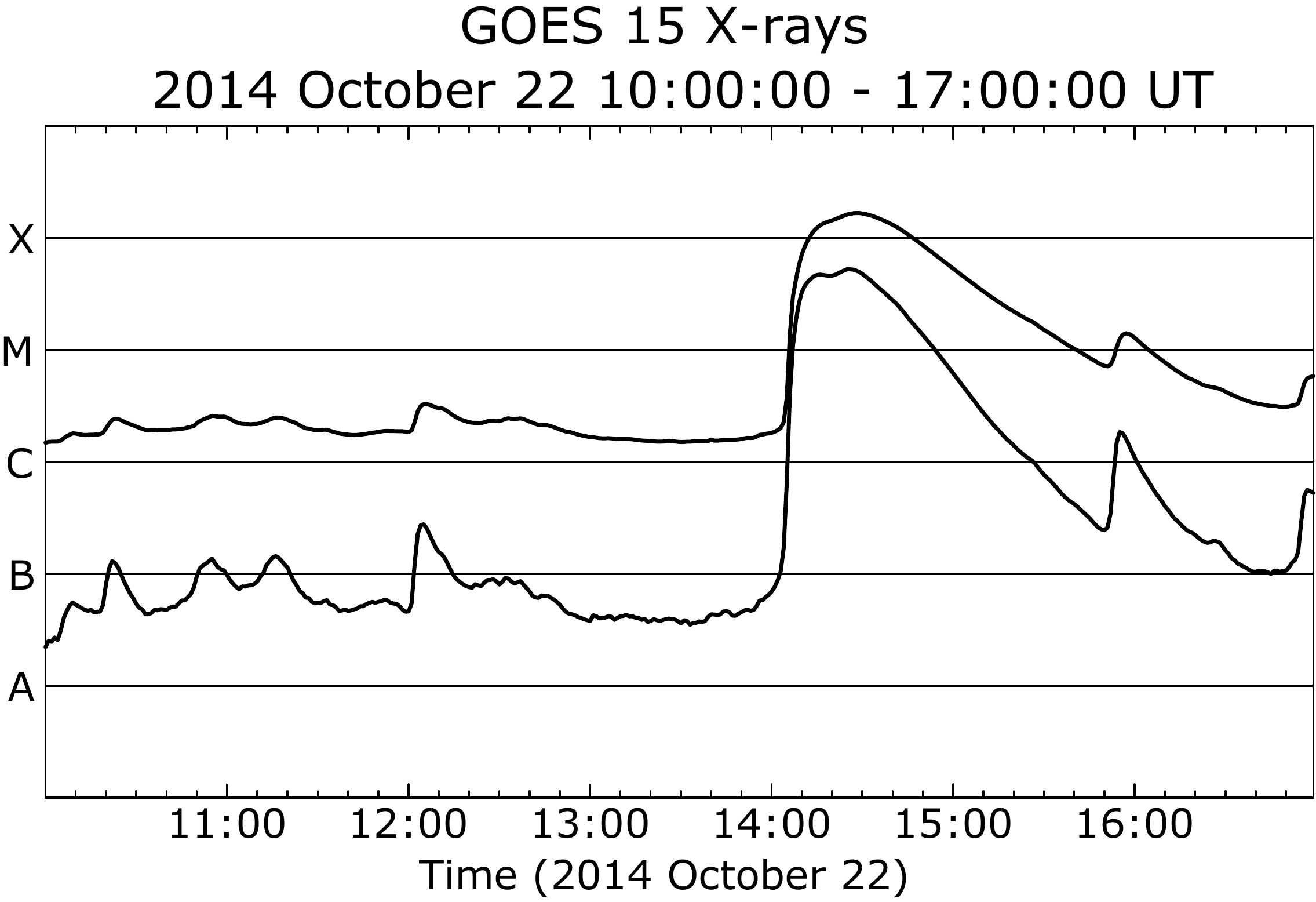}
\caption{
The soft X-ray light curve observed by GOES from 10:00-17:00 UT on 2014 October 22.
The upper and lower curves indicate 1-8 {\AA} and 0.5-4 {\AA} channels, respectively.
The start, peak, and the end time of the X1.6 flare are 14:02 UT, 14:28 UT, and 14:50 UT, respectively.
The C3.2 event also occurred on 12:00 UT at the same region.
The M1.4 event that start at 15:51 UT and following the X1.6 flare occurred in different region (on the southeast limb).
}
\label{fig:goes}
\end{figure*}

The AR NOAA 12192 appeared at the east limb on 2014 October 16.
It was the biggest sunspot region in the solar cycle 24, and it produced more than 130 flares that are larger than C1.0 class.
The X1.6 flare, that is the second X-class flare of the six X-class flares occurred in the AR, near the disk center (S$14^{\circ}$, E$13^{\circ}$) on October 22.
Figure~\ref{fig:goes} shows the soft X-ray light curve observed by the Geostationary Operational Environmental Satellite (GOES) 1-8 {\AA} and 0.5-4 {\AA} channels, and the start, peak, and the end time of the X1.6 flare are 14:02 UT, 14:28 UT, and 14:50 UT, respectively.
The C3.2 flare occurred on 12:00 UT prior to the X1.6 flare at the same region.
There was no flaring activities after the C3.2 flare, as represented by the flat X-ray light curve in the period of 12:30-14:00 UT, but the X1.6 flare suddenly happened.
Moreover, there is no coronal mass ejection (CME) and solar proton event (SPE) related to the X1.6 and following C3.2 flares \footnote{We checked existence of CME and SPE from SOHO LASCO CME catalog (\url{https://cdaw.gsfc.nasa.gov/CME_list/}) of the CDAW Data Center by NASA and SPE event list (\url{https://umbra.nascom.nasa.gov/SEP/}) provided by NOAA Space Environment Services Center, respectively.}.

We analyzed magnetograms and filtergrams obtained by SDO/Helioseismic and Magnetic Imager (HMI; \citet{schou12}) and Atmospheric Imaging Assembly (AIA; \citet{lemen12}).
We used HMI LOS/vector magnetograms in \ion{Fe}{1} line (6173 {\AA}) to investigate photospheric magnetic field structures.
Also, we used AIA filtergrams in 1600 {\AA} (continuum and \ion{C}{4} line) and 131 {\AA} (\ion{Fe}{8}, \ion{}{21}), which are sensitive to emission from the transition region and upper chromosphere, and the transition region and flaring corona, respectively \citep{lemen12}.
The cadence for LOS magnetograms, vector magnetograms, and AIA 1600 {\AA}/131 {\AA} images were $45~sec.$, $12~min.$, and $24~sec.$/$12~sec.$, respectively.
We summarized observation parameters for SDO/HMI and AIA in Table~\ref{table:SDO}.

IRIS successfully scanned over the central region of the AR with the slit that is tilted to $45~degree$ from the solar NS direction, and took spectra and slit-jaw images (SJIs).
We used the spectra and SJIs in the period of 10:00 UT (four hours before the flare onset) to 16:00 UT (two hours after the flare onset).
The spectra were obtained with coarse raster scan, and used \ion{C}{2} (1330 {\AA}), \ion{Si}{4} (1400 {\AA}), and \ion{Mg}{2} k (2796 {\AA}) lines in this study.
These were taken at eight positions of 0.33$^{\prime\prime}$ slit every $16.4~sec.$, i.e. it took approximately $2~min.$ to scan 8 steps (14$^{\prime\prime}$ $\times$ 174$^{\prime\prime}$ FOV).
The spectral resolution is 50 m{\AA} and 80 m{\AA}, and the pixel resolution is 0.33$^{\prime\prime}$ and 0.4$^{\prime\prime}$ for far-ultraviolet (FUV; \ion{C}{2} and \ion{Si}{4}) and near-ultraviolet (NUV; \ion{Mg}{2} k) images, respectively.
The SJIs were obtained in \ion{C}{2} (1330 {\AA}) and \ion{Mg}{2} k (2796 {\AA}) lines every $32.7~sec.$ with a FOV of 167$^{\prime\prime}$ $\times$ 174$^{\prime\prime}$.
The spatial resolution is 0.33$^{\prime\prime}$ and 0.4$^{\prime\prime}$ for \ion{C}{2} and \ion{Mg}{2} k images, respectively.
We summarized observation parameters in Table~\ref{table:IRIS_EIS}.

We also analyzed sparse raster scan data obtained by Hinode/EIS from 13:01:56 UT to 15:56:56 UT on 2014 October 22.
EIS observed the AR with 12 spectral windows.
In this study we focused on three strong emission lines, \ion{Fe}{12}, \ion{Fe}{15}, and \ion{He}{2}, which are sensitive to the temperature of log T = 6.2, 6.3, and 4.9, respectively.
The spectra were taken at 20 positions of 2$^{\prime\prime}$ slit every $9~sec.$, and it took $3~min.$ to scan 20 steps (i.e. 59$^{\prime\prime}$ $\times$ 152$^{\prime\prime}$ FOV).
The spectral resolution is 22 m{\AA} as summarized in Table~\ref{table:IRIS_EIS}.

%%%ANALYSIS METHODS%%%
\section{Analysis Methods} \label{sec:analysis}

\subsection{Analysis for the Imaging Data} \label{sec:analysis_imaging}

We used the analysis method of \citet{bamba13, bamba14} to co-align HMI LOS magnetograms and AIA images.
Here, we briefly review the method.
We used HMI level 1.5 LOS magnetograms ({\tt hmi.M\_45s} series) and AIA level 1.0 data ({\tt aia.lev1\_uv\_24s} and {\tt aia.lev1\_euv\_12s} series) from 10:00 UT to 17:00 UT on 2014 October 22.
We first calibrated the HMI LOS magnetograms and AIA images using the {\tt aia\_prep} procedure in Solar Soft Ware (SSW).
Through this process, we reduced the spatial fluctuations, and rotated the images so that the solar EW and NS axes are aligned with the horizontal and vertical axes of the image, respectively.
Moreover, we resampled LOS magnetograms and AIA images to the same size because the pixel scales are different between HMI and AIA.
Thus the positions of the LOS magnetograms and AIA images were aligned.
Next, we chose a LOS magnetogram and an AIA image closest in time, and these two images were superimposed onto each other.
Figure~\ref{fig:HMI}(a-c) shows the LOS magnetograms on which the magnetic polarity inversion lines (PILs; lines of $0G$, green), and the contour of the strong brightening in AIA 1600 {\AA} (red) are over plotted.
%We investigated the spatial and temporal correlation between the photospheric LOS magnetic field structures and the strong brightenings in the upper chromosphere and the transition region.
AIA 131 {\AA} images, on which the PIL of HMI LOS magnetogram is partially overlaid, are also shown in Figure~\ref{fig:AIA131}.

We also investigated the distribution of the relative shear angle $\chi$ over the AR before the flare onset, using the analysis method proposed by \citet{bamba_inpress}.
They developed the method for Hinode/SP data, but here we applied the method to HMI vector magnetogram, i.e. the Spaceweather HMI Active Region Patch (SHARP: {\tt hmi.sharp\_cea\_720s} series), because there were no SP data before/after six hours from the onset time of the X1.6 flare.
The magnetic field of SHARP data series has been calibrated assuming the Milne-Eddington atmosphere and remapped to a Lambert Cylindrical Equal-Area projection, and its $180^{\circ}$ ambiguity in the horizontal component is resolved using a minimum energy method (\citet{metcalf94, leka09, leka12}).
Therefore, we got the horizontal and radial component of magnetic field (${\bm B_{x}}$, ${\bm B_{y}}$, ${\bm B_{z}}$) from the SHARP data.
In this study, we used the SHARP vector magnetogram taken at 13:35 UT on 2014 October 22, when is the closest in time to the last precursor brightening in AIA 1600 {\AA} images over a distinctive-shape PIL.
We calculated the potential field ${\bm B_{p}}$ using the {\tt fff} procedure in the {\tt nlfff} package (developed by Dr. Yuhong Fan) in SSW.
Then we measured the angles between the potential field vector ${\bm B_{p}}$ and the horizontal field vector $\bm B_{h} = \sqrt{{B_{x}}^2 + {B_{y}}^2}$ in each pixel.
In this paper, we call them ``relative shear angle $\chi$".
The relative shear angle $\chi$ is defined between $\pm180^{\circ}$, where $0^{\circ}$ means vectors ${\bm B_{p}}$ and ${\bm B_{h}}$ are oriented in the same direction.
The direction of vector ${\bm B_{h}}$ deviates from vector ${\bm B_{p}}$ to counter-clockwise (clockwise) when the magnetic helicity is positive (negative) and the value of $\chi$ is positive (negative).
We colorized the relative shear angles as shown in Section~\ref{sec:results_SDO} and Figure~\ref{fig:HMI}(d).

With regards to IRIS data, we used level 2 data in which dark-current subtraction, flat fielding, and geometrical correction are taken into account.
Because AIA 1600 {\AA} images and HMI LOS magnetograms are already co-aligned using the above mentioned procedure, we can easily co-align between  HMI LOS magnetograms and IRIS SJIs via AIA 1600 {\AA} images.
We drew the PILs of HMI LOS magnetograms onto IRIS SJIs in order to check the spatial and temporal correlation between the precursor brightening and magnetic field (Figure~\ref{fig:velocity}(a, b)).

\subsection{Analysis for the Spectrum Data} \label{sec:analysis_spectra}

We calculated the Doppler velocities for \ion{Si}{4} line, using IRIS coarse scan data.
First, we performed wavelength calibration using a photospheric line (S I, 1401 {\AA}).
We decided the central value $\lambda_{SI\_obs}$ of the observed wavelength for S I line by the single Gaussian fitting.
Here, we integrated the observed line profile over the region, where was relatively quiet and out of the sunspot umbra.
We applied the single Gaussian fitting to the integrated line profile and repeated the process for the time period of 11:00 UT to 12:00 UT on October 22.
We averaged the fitted line center value over the time period of 11:00 UT to 12:00 UT, and obtained the observed line center $\lambda_{SI\_obs} = 1401.5583$ ({\AA}).
Then we got {\it dw} by subtracting $\lambda_{SI\_obs}$ from $\lambda_{SI\_lab} = 1401.5136$ ({\AA}), which is the laboratory wavelength of S I line \citep{kaufman93}.
\begin{displaymath}
dw = \lambda_{S I\_lab} -  \lambda_{S I\_obs}
\end{displaymath}
Using  {\it dw}, we calibrated \ion{Si}{4} line as 
\begin{displaymath}
\lambda_{0, Si IV} = \lambda_{Si IV\_lab} - {\it dw},
\end{displaymath}
where $\lambda_{Si IV\_lab}$ is the laboratory wavelength of \ion{Si}{4} line (1402.769 {\AA}; \citet{brekke97}), 
%from the CHIANTI line list \footnote{The CHIANTI is an atomic database for spectroscopic diagnostics of astrophysical plasmas developed by George Mason Univ., Univ. of Michigan, and Univ. of Cambridge (\url{http://www.chiantidatabase.org/chianti.html}). In this study, we chose the theoretical wavelength for S I, \ion{Si}{4}, \ion{C}{2}, and \ion{Mg}{2} k lines from the spectral line list of the CHIANTI version 7.0.}; \citep{landi12}, 
and $\lambda_{0, Si IV}$ is the reference wavelength of \ion{Si}{4} in the IRIS raster scan observation.

Next we identified the center of observed \ion{Si}{4} line $\lambda_{Si IV\_obs}$ using {\tt eis\_auto\_fit} procedure in SSW package, which apply the single Gaussian fitting to line profiles.
Then we calculated the Doppler velocities as
\begin{displaymath}
v = \frac{\Delta\lambda_{Si IV}c}{\lambda_{0, Si IV}},
\end{displaymath}
where $\Delta\lambda_{Si IV} = \lambda_{0, Si IV} - \lambda_{Si IV\_obs}$ and $c$ is the speed of light.
We measured the Doppler velocities from 10:00 UT to 16:00 UT on October 22, and made maps of the Doppler velocities, intensities, line width for \ion{Si}{4} line using {\tt eis\_get\_fitdata} procedure (as can be seen in Figure~\ref{fig:velocity}(c-f)).
Using the Si IV maps of Figure~\ref{fig:velocity}(c-f), we detected the times and locations when/where the line profiles were distinctive such as blueshifted or redshifted.

\begin{figure*}
\epsscale{2.20}
\plotone{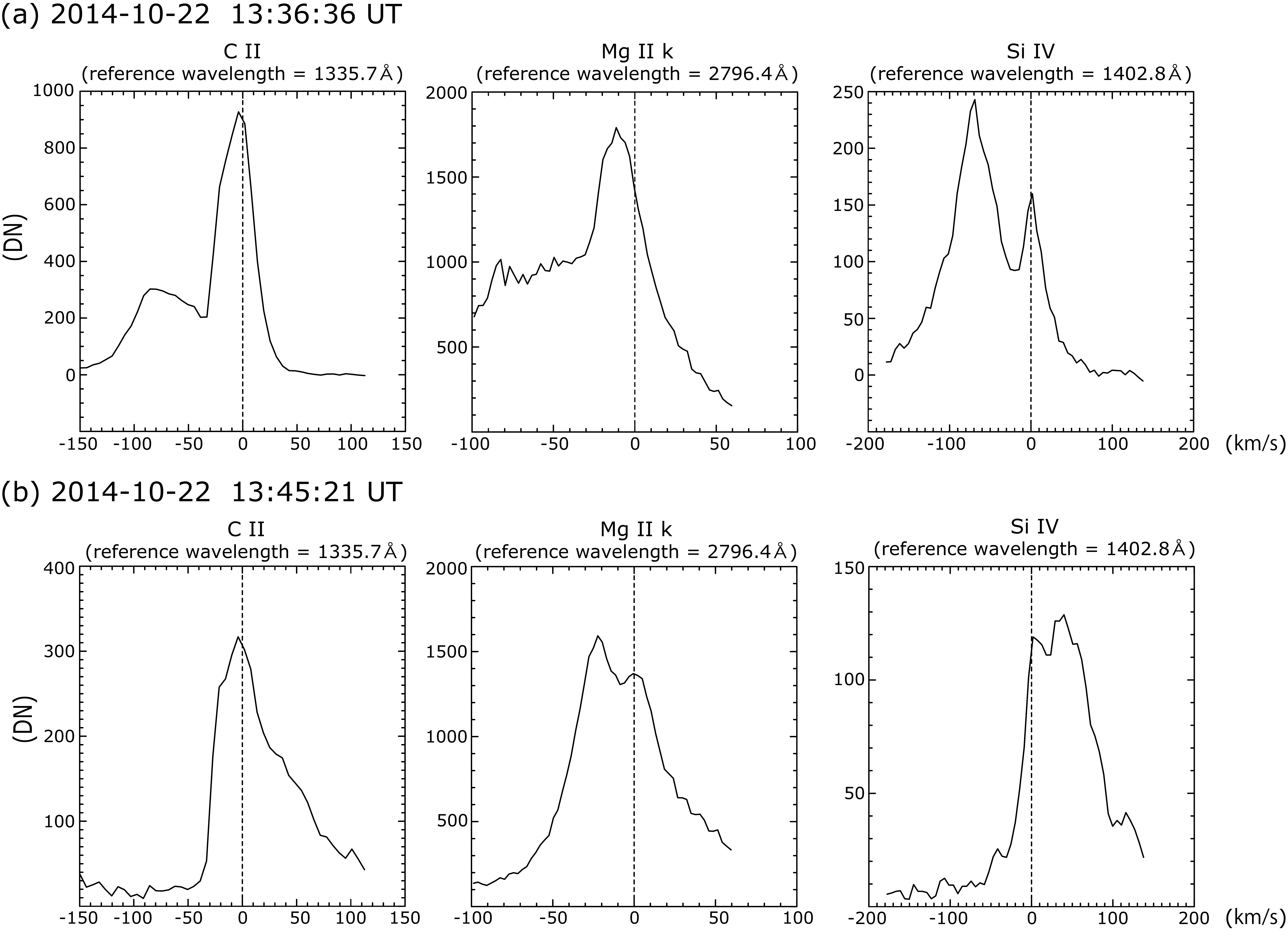}
\caption{
The line profiles of \ion{C}{2}, \ion{Mg}{2} k, and \ion{Si}{4} lines from a single slit spatial pixel where is closest to the center of TR1 on the slit position 4.
The vertical broken lines indicate the zero velocity which is defined as the reference wavelength of each line.
(a) Blueshift profiles taken at 13:36:36 UT at the fourth slit position (solid lines).
(b) Redshift profiles taken at 13:45:21 UT at the fourth slit position (solid lines).
}
\label{fig:lineprofiles}
\end{figure*}

We displayed the spectrum images and line profiles of \ion{Si}{4}, \ion{C}{2}, and \ion{Mg}{2} k lines at the distinctive time and the location (Figures~\ref{fig:spectra} and \ref{fig:lineprofiles}).
%With regard to \ion{C}{2} and \ion{Mg}{2} k lines, we displayed the spectrum images and line profiles, which were converted from wavelength ({\AA}) to velocity ($km~s^{-1}$).
%Basically, we applied the wavelength calibration also for \ion{C}{2} and \ion{Mg}{2} k lines using S I line as explained in the previous paragraph, and we obtained the reference wavelengths as
%\begin{displaymath}
%\lambda_{0, C II} = \lambda_{C II\_theo} - {\it dw},
%\end{displaymath}
%\begin{displaymath}
%\lambda_{0, Mg II k} = \lambda_{Mg II k\_theo} - {\it dw}.
%\end{displaymath}
In these images and plots, the wavelength ({\AA}) was converted to velocity ($km~s^{-1}$), and we defined the reference wavelength $\lambda_{0, Si IV}$ (1402.8 {\AA}), $\lambda_{0, C II}$ (1335.7 {\AA}), and $\lambda_{0, Mg II k}$ (2796.4 {\AA}) from Table 4 of \citet{depontieu14} as $0~km~s^{-1}$.
The line profiles in Figure~\ref{fig:lineprofiles} show the intensities of \ion{Si}{4}, \ion{C}{2}, and \ion{Mg}{2} k lines at the time and location when/where the significant blueshift and redshift are seen in Figure~\ref{fig:velocity}(c-f).
Using these spectrum images and line profiles, we discussed the Doppler velocities for \ion{Si}{4}, \ion{C}{2}, and \ion{Mg}{2} k lines, with order of approximately $10~km~s^{-1}$  accuracy, in Sections~\ref{sec:results_IRIS_EIS} and \ref{sec:discussion}.

\begin{figure*}
\epsscale{2.30}
\plotone{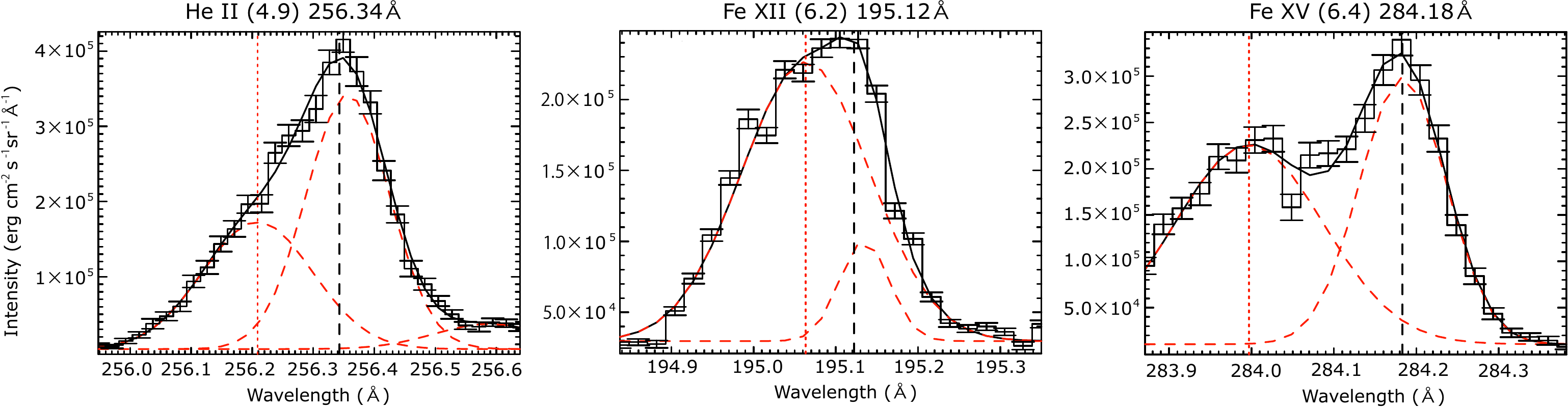}
\caption{
The line profiles of \ion{He}{2}, \ion{Fe}{12}, and \ion{Fe}{15} lines taken by Hinode/EIS.
The profiles were averaged over a period around 13:34:40 UT over the region of the slit length in TR1.
The red broken lines represent each Gaussian component, and the red dashed vertical lines represent the line center of the fitted line profiles for blueshifted component.
The black solid lines are the fitted line profile, and the vertical broken lines indicate the reference wavelength which is corresponding to the zero velocity.
All the profiles show significant blueshift.
}
\label{fig:EIS_profile}
\end{figure*}

EIS data from the raster are processed using the EIS team provided software ({\tt eis\_prep}), which corrects for the flat field, dark current, cosmic rays, hot pixels, and slit tilt.
For thermal reasons, there is an orbital variation of the line position causing an artificial Doppler shift of $\pm$20 km s$^{-1}$ which follows a sinusoidal behavior.
This orbital variation and wavelength calibration of the line position were corrected using the house keeping data \citep{kamio10}.
We measured the average line centers of \ion{He}{2}, \ion{Fe}{12}, and \ion{Fe}{15} lines before the flare (between 13:01-13:51 UT) for the quieter region of the EIS FOV, in order to obtain the reference wavelengths of 256.343 {\AA} (\ion{He}{2}), 195.121 {\AA} (\ion{Fe}{12}), and 284.183 {\AA} (\ion{Fe}{15}), respectively.
After above wavelength calibration, we can determine the LOS velocities estimated by Doppler shift in \ion{He}{2}, \ion{Fe}{12}, and \ion{Fe}{15} emission lines with a few $km~s^{-1}$ uncertainty.
Note that the Doppler velocities in Figure~\ref{fig:EIS} are derived by applying the single Gaussian fitting, while we show line profiles with the double Gaussian fitting in Figure~\ref{fig:EIS_profile}.

%%%RESULTS%%%
\section{Results} \label{sec:results}

\subsection{Overview of the Flare} \label{sec:results_SDO}

Figure~\ref{fig:HMI}(a-c) shows the LOS magnetograms observed by SDO/HMI.
White/black indicates positive/negative polarity of LOS magnetic field, and green lines and the red contours outline the PILs ($0~G$ line of the LOS magnetic field) and the strong brightening in AIA 1600 {\AA} images, respectively.
In panel (a), two isolated positive polarity regions TR1 and TR2 are marked by the yellow circles, and precursor brightening is seen over the southwest PIL of both TR1 and TR2.
TR1 locates in the negative sunspot (NS) and it has emerged from October 19.
The precursor brightening is seen over the southwest PIL, and it starts from 12:00 UT and strengthen around 13:40 UT on October 22.
TR2 has also emerged as an isolated positive polarity region from October 21, but it gradually merged into the positive sunspot (PS).
Small brightening is intermittently seen over the southwest PIL of TR2 from 10:00 UT, and it continues until $3~min.$ before the flare start time.
On the other hand, the brightening in TR1 finishes approximately $10~min$ earlier than the flare start time.
The initial flare ribbons appear as indicated by the yellow arrows in panel (b), and these are developed to typical two flare ribbons which are clearly seen in panel (c).

We also quantitatively investigated the distribution of the relative shear angle $\chi$, that is defined as the angles between the potential field vector ${\bm B_{p}}$ and horizontal field vector ${\bm B_{h}}$, using a HMI vector magnetogram.
Figure~\ref{fig:HMI}(d) shows the distribution of $\chi$ at 13:35 UT, when the strong brightening is seen over TR1.
The red/blue corresponds to positive/negative values of $\chi$, and the black lines indicate the PILs ($0~G$ line of the radial component ${\bm B_{z}}$ of the magnetic field).
The angle $\chi$ is around $-90^{\circ}$ (blue) along the flaring PIL while it is around $90^{\circ}$ (red) at the southwest side of both TR1 and TR2.
Therefore, the distribution of the relative shear angle $\chi$ suggests that the local magnetic shear in the southwest side of both TR1 and TR2 towards the opposite direction from the global magnetic shear along the flaring PIL between PS-NS.

Figure~\ref{fig:AIA131}(b-f) shows the time series of AIA 131 {\AA} images.
The small images in panel (c) are enlarged view of the region where is surrounded the red broken line in panel (b).
In panels (g, h), we plotted the total intensity ($DN$ unit) in the region where is surrounded by the red rectangle in panel (b).
The intensity enhancement around 12:00 UT in panel (g) is corresponding to the C3.2 flare.
We observed that the coronal loop connecting PS-NS bright before the C3.2 event, although the intensity become low and quiet after the C-flare.
Especially after 13:00 UT, i.e. from one hour before the X-flare, the coronal loop became dark as seen in panel (b).
A jet-like small eruption from TR1 to southwestward is observed at 13:35:00-13:36:00 UT (about $30~min.$ before the X-flare onset), then the strong precursor brightening is observed, as indicated by the yellow arrows in panels (c, d).
Moreover, the coronal loop connecting PS-NS in the region where is surrounded by the red square in panel (b) become bright after the precursor brightening (compare panels (b, d)).
It is clearly seen in panel (h) in which the baseline of the intensity is enhanced after the precursor brightening in TR1 that marked by the blue vertical line.
%The {\bf precursor} brightening over TR1 continues until the X1.6 flare onset whereas the brightening over TR2 blink and then disappear on 13:58 UT.
Then two flare ribbons appear at the foot points of the coronal loop connecting PS-NS (panel (f)).

\subsection{Result from Spectroscopic Data} \label{sec:results_IRIS_EIS}

Figure~\ref{fig:velocity}(a, b) shows IRIS SJIs at 13:36 UT (about $30~min.$ before the flare onset).
We overlaid the PILs of HMI LOS magnetogram with green line.
The eight slit positions for spectroscopic observation are indicated by gray and blue lines.
IRIS successfully scanned over TR1, and we observed strong blueshift signals which are simultaneously seen with the last precursor brightening in southwest of TR1.
Figure~\ref{fig:velocity}(c-f) shows the distributions of the intensity, Doppler velocity, and line width for \ion{Si}{4} line.
%To understand the plasma motions during the {\bf precursor} activity, we have fitted line profiles by a single Gaussian.
The brightening in TR1, such as seen in Figure~\ref{fig:HMI}(a) is marked by the yellow circle in each intensity image (left columns).
There is no significant Doppler signal and line broadening over TR1 although transient brightening is seen in the region at 13:28:58 UT (Figure~\ref{fig:velocity}(c)).
However, significant blueshift is observed just after the last precursor brightening appeared in the southwest region of TR1 at 13:35:31 UT, as seen in panel (d).
The blueshift is weakened in the brief period of $2~min.$ (panel (e)), and further $7~min.$ later, redshift signal is observed in the same region (panel (f)).
The averaged velocity of the blueshift in panel (d) and the redshift in panel (f) are $\sim-50~km~s^{-1}$ and $\sim25~km~s^{-1}$ estimated by a single Gaussian fitting, respectively.
Line broadenings are continuously observed during the last precursor brightening is seen in panels (d-f).

Figure~\ref{fig:spectra} displays the spectral images of \ion{Si}{4}, \ion{C}{2}, and \ion{Mg}{2} k lines.
These spectra were taken at a single slit spatial pixel in the fourth slit position (indicated by the blue line in Figure~\ref{fig:velocity}(a, b)) and just over the region where the strong brightening is seen at 13:35:31 UT in Figures~\ref{fig:HMI}(a), \ref{fig:AIA131}(c), and \ref{fig:velocity}(a, b).
%We define the reference wavelengths $\lambda_{0, Si IV}$, $\lambda_{0, C II}$, and $\lambda_{0, Mg II k}$ as $0~km~s^{-1}$.
The spectra clearly shows blueshift as indicated by the yellow arrows, and the locations of the blueshift are consistent to the last precursor brightening over TR1 around 13:36 UT.
The velocities are more than $-50~km~s^{-1}$ (almost up to $-100~km~s^{-1}$) in each lines which represent from the chromosphere to transition region temperature plasma.
The blueshift was observed in all the three lines in Figure~\ref{fig:spectra}.
%Note that all three lines show the almost same velocity.
%This indicates that the Doppler shifts which is found during the precursor activity might be temperature independent flow.

Figure~\ref{fig:lineprofiles} is the line profiles of the blueshift and redshift taken at a single slit spatial pixel in the fourth slit position.
The solid lines are the profiles at the point where the blueshift and redshift are seen in TR1, at 13:36:36 UT in panels (a) and at 13:45:21 UT in panel (b).
The horizontal axes are converted from the wavelength ({\AA}) to the velocity ($km~s^{-1}$), and the zero velocity is indicated by the vertical broken lines.
%The reference wavelengths $\lambda_{0, Si IV}$, $\lambda_{0, C II}$, and $\lambda_{0, Mg II k}$ are defined as $0~km~s^{-1}$, and the zero velocity is indicated by the vertical broken lines.
In panel (a), all the spectral line profiles are clearly blueshifted.
We can see double intensity peak in the \ion{Si}{4} line profile although \ion{Si}{4} usually shows single intensity peak profile, and the velocity is almost $-70~km~s^{-1}$ at that time at the point.
\ion{C}{2} and \ion{Mg}{2} k lines usually have double intensity peak, and the reference wavelength should be the center of the absorption part in the middle of two emission part according to our definition in Section~\ref{sec:analysis_spectra}.
However, the center of the absorption part is blueshift in \ion{C}{2} line, and the velocity is $\sim40~km~s^{-1}$ at that time at the point.
In case of \ion{Mg}{2} k line, it is difficult to determine the velocity because the line profile shapes much different from that in a quiet region.
Nevertheless, it obviously has blueshifted component.
In panel (b), the significant redshift can be seen in \ion{Si}{4} and \ion{C}{2} lines.
Especially, \ion{Si}{4} line profile shows clear two intensity peak and the velocity of redshifted part is almost $40~km~s^{-1}$.
\ion{Mg}{2} k line shows clear double intensity peak, but Doppler shift is not significant with the order of velocity measurement ($\sim 10~km~s^{-1}$) in the present study.
Thus, the redshifts observed just after precursory activity at all the temperature.
%Thus, the redshifts observed just after precursory activity might temperature dependent flows.
%It can be interpreted that the plasma could not intrude into the temperature region which \ion{Mg}{2} k line is sensitive ($log(T)\sim4.0$) because the red-shifted velocity of plasma is slower than that of blue-shifted plasma.

\begin{figure*}
\epsscale{2.00}
\plotone{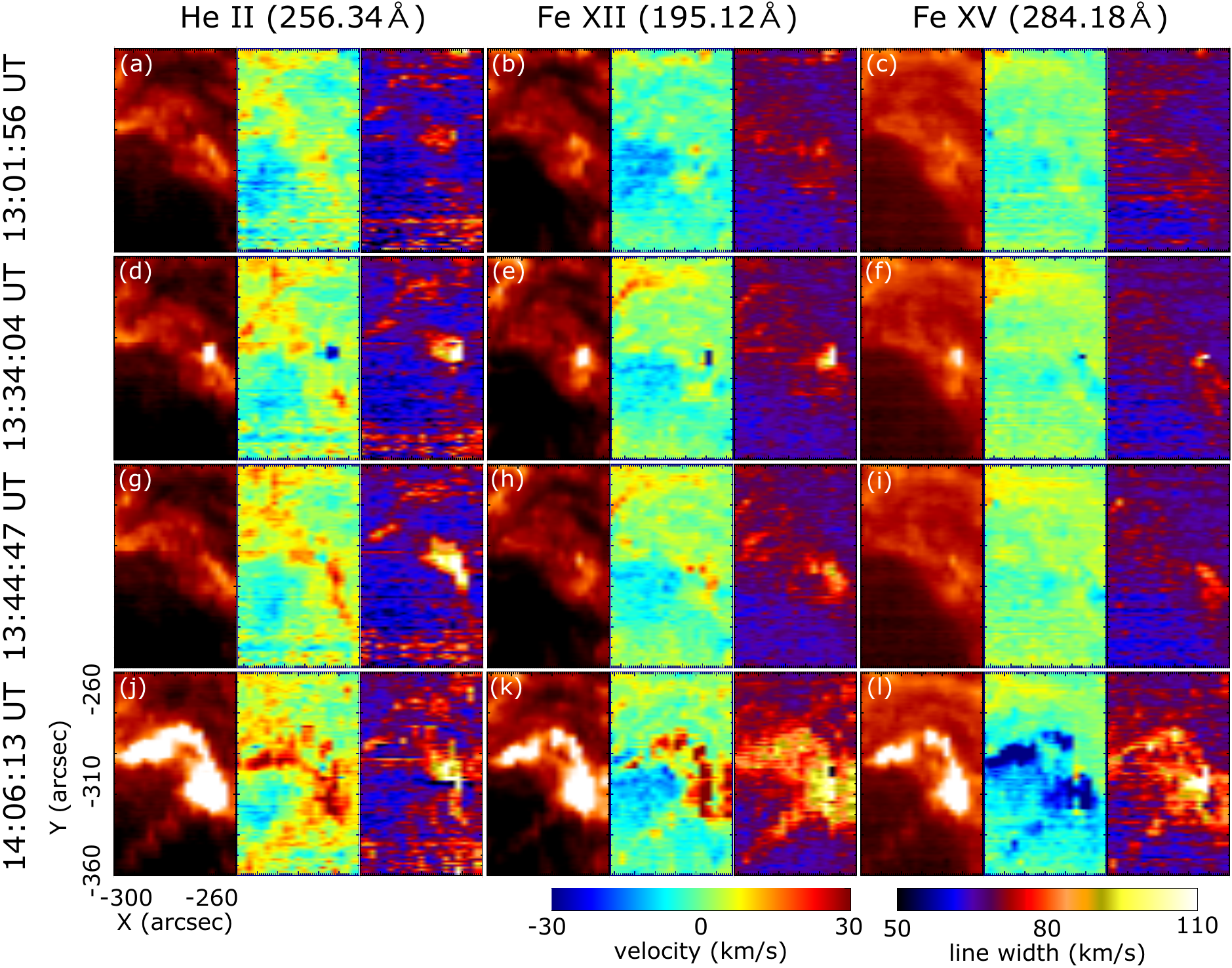}
\caption{
Intensities (left), Doppler velocities (middle), and line width (right) for \ion{He}{2}, \ion{Fe}{12}, and \ion{Fe}{15} lines obtained by Hinode/EIS.
%EIS scans the FOV from west to east (right to left in the images), and it takes about $3~min.$ to scan the FOV.
The start times of scanning are shown in left column, and color bars for Doppler velocity and line width are in right bottom.
(a-c) There is no significant brightening, Doppler signal, and line broadening in the period when GOES X-ray and AIA 131 {\AA} intensity light curves are flat.
(d-f) Strong blueshift signals are seen with the precursor brightening over TR1 around 13:36 UT. Strong line broadenings are also seen especially in \ion{He}{2} and \ion{Fe}{12} lines.
(g-i) Weak redshift signals are seen only in \ion{He}{2} and \ion{Fe}{12} lines in the region where blueshift signal were observed. In \ion{Fe}{15} line, there are no significant Doppler signal over TR1.
(j-l) Flare kernel in the negative sunspot NS is redshifted in \ion{He}{2} and \ion{Fe}{12} lines while it is blueshifted in \ion{Fe}{15} line.
}
\label{fig:EIS}
\end{figure*}

Figure~\ref{fig:EIS} shows the similar plot to Figure~\ref{fig:velocity}(c-f): the distributions of the intensity, Doppler velocity, and line width for \ion{He}{2}, \ion{Fe}{12}, and \ion{Fe}{15} lines obtained by Hinode/EIS.
%The velocities and line widths are estimated by a single Gaussian fitting.
The FOV covers the following negative sunspot region NS in Figure~\ref{fig:HMI}, and the dark region in left bottom of each intensity maps in Figure~\ref{fig:EIS} corresponds to the sunspot umbra.
There is no significant brightening, Doppler signal, and line broadening at 13:01:36 UT when the AIA 131 {\AA} intensity is constantly low as seen in Figure~\ref{fig:AIA131}(h).
Conversely, we can clearly see the strong brightenings, strong blueshifts, and significant line broadenings at 13:34:04 UT in Figure~\ref{fig:EIS}(d-f).
These are corresponding to the last precursor brightening, blueshift, and line broadening seen in Figure~\ref{fig:HMI}(a), \ref{fig:AIA131}(c), and \ref{fig:velocity}(d).

Figure~\ref{fig:EIS_profile} shows the line profiles of \ion{He}{2}, \ion{Fe}{12}, and \ion{Fe}{15} lines corresponding to the blueshift signals in Figure~\ref{fig:EIS}(d-f).
The line profiles were averaged over a period around 13:34:40 UT over the region of the slit length in TR1.
The vertical black broken lines indicate the reference wavelengths.
The red lines show the double Gaussian fitting results.
The blueshifted components are almost up to $100~km~s^{-1}$ for \ion{He}{2} and \ion{Fe}{12}, although the line profile in \ion{Fe}{15} shows much faster than the others.
There are \ion{Fe}{17} (283.95 ~\AA) and \ion{Al}{9} (284.03 ~\AA) in the spectral window.
These blending lines might affect the double Gaussian fitting.
The redshifts are also seen at 13:44:47 UT in \ion{He}{2} and \ion{Fe}{12} lines  of Figure~\ref{fig:EIS}(g, h), as well as seen in \ion{Si}{4} line of Figure~\ref{fig:velocity}(f).
The significant line broadenings are still remaining not only \ion{He}{2} and \ion{Fe}{12} lines but also \ion{Fe}{15} line, although there is no significant Doppler signal in Figure~\ref{fig:EIS}(i). 
The flare ribbon in the negative polarity region NS can be seen at 14:06:13 UT in Figure~\ref{fig:EIS}(j-l).
We can clearly observe the typical temperature dependent flows which correspond to the chromospheric evaporation \citep{milligan2009, imada15}.
The detail analysis for the chromospheric evaporation in this flare has been done by \cite{lee_submitted}.

\subsection{Summary of the Results} \label{sec:results_summary}

\begin{enumerate}
\item Precursor brightening is observed in the southwest region of the small isolated positive regions TR1 and TR2, which are located between PS-NS. Especially, the last precursor brightening near TR1 around 13:36 UT is significant, and a jet-like small eruption from TR1 to the southwestward is approximately simultaneously observed. After the last precursor brightenning, the coronal loop connecting PS-NS become brighter to the flare onset.
\item The southwest region of both TR1 and TR2 locally has the magnetic shear that towards opposite direction to the global shear along the flaring PIL between PS-NS. In other words, the magnetic helicity is locally positive in the southwest of both TR1 and TR2, while the helicity along the flaring PIL is negative.
\item 
Strong blueshift (up to $100~km~s^{-1}$) is detected in the southwest region of TR1 simultaneously with the last precursor brightening around 13:36 UT. It is seen in all the lines (in all the temparetures) that analyzed in this study (\ion{Si}{4}, \ion{C}{2}, and \ion{Mg}{2} k lines obtained by IRIS and \ion{He}{2}, \ion{Fe}{12}, and \ion{Fe}{15} lines obtained by Hinode/EIS).% and it seems that the these flows do not depend on the temperature.
The redshift is detected in the same location around 13:44 UT; later than the bluehisft in \ion{Si}{4} and \ion{C}{2} lines of IRIS.
\end{enumerate}

%%%DISCUSSION%%%
\section{Discussion} \label{sec:discussion}

Here we discuss the triggering scenario of the X1.6 flare.
As we summarized in Section~\ref{sec:results_summary}, the last precursor brightening and the small jet-like eruption are observed corresponding to TR1, and then the large-scale coronal loop connecting PS-NS become bright.
It suggests that a precursor activity, that is likely a small magnetic reconnection and corresponding jet, in TR1 makes global structure in the corona unstable.
TR1 is an opposite polarity emerging flux that has been proposed as a candidate of a ``flare-triggering structure'' by some theoretical models (e.g. \citet{heyvaerts77, chenshibata00, kusano12}).

\begin{figure*}
\epsscale{2.30}
\plotone{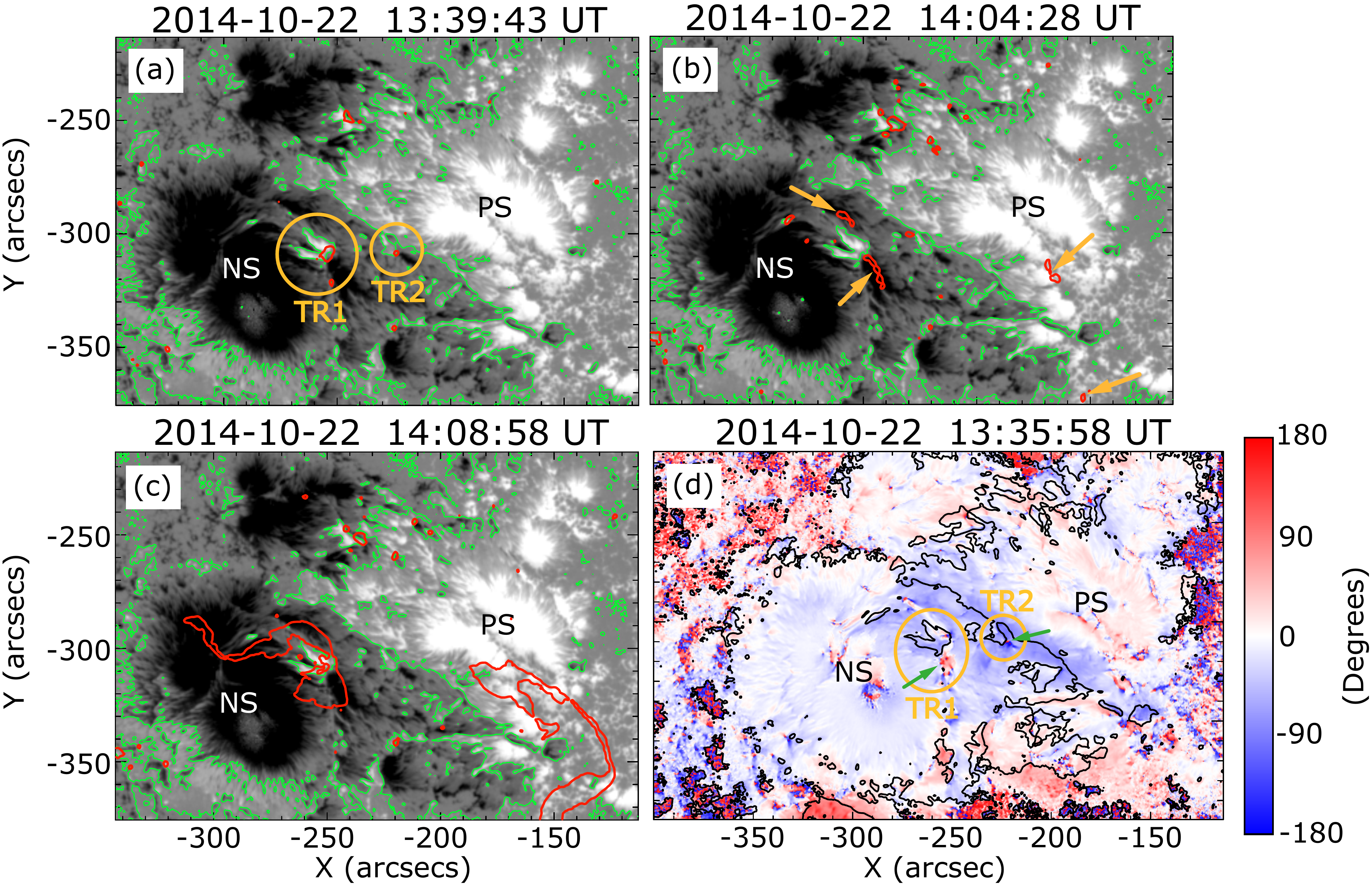}
\caption{
(a)-(c) Temporal evolution of LOS magnetic field and precursor brightening seen in AIA 1600 {\AA}.
The white/black indicates positive/negative polarity of the LOS magnetic field, and the intensity scale is $\pm~1000~G$.
Green lines indicate the PILs ($0~G$ line of the LOS magnetic field), and the red contours outline the strong brightening in AIA 1600 {\AA} images with an intensity of $2000~DN$.
TR1 and TR2, the region where the strong brightening is seen, are pointed out by the yellow circles in panel (a).
The initial flare ribbons that appears as four bright points are indicated by the yellow arrows in panel (b).
(d) Distribution of the relative shear angle $\chi$ which is defined as the angles between the potential field vector ${\bm B_{p}}$ and horizontal field vector ${\bm B_{h}}$.
The black lines are the PILs ($0~G$ line of the radial component ${\bm B_{z}}$ of the magnetic field), red/blue corresponds to positive/negative values of $\chi$, i.e. the magnetic helicity.
TR1 and TR2, the region where the strong emissions were seen, is pointed out by the yellow circles.
The regions where are indicated by the green arrows are locally have reversed magnetic shear (i.e. opposite magnetic helicity) to the global magnetic shear along the flaring PIL.
}
\label{fig:HMI}
\end{figure*}

\begin{figure*}
\epsscale{2.30}
\plotone{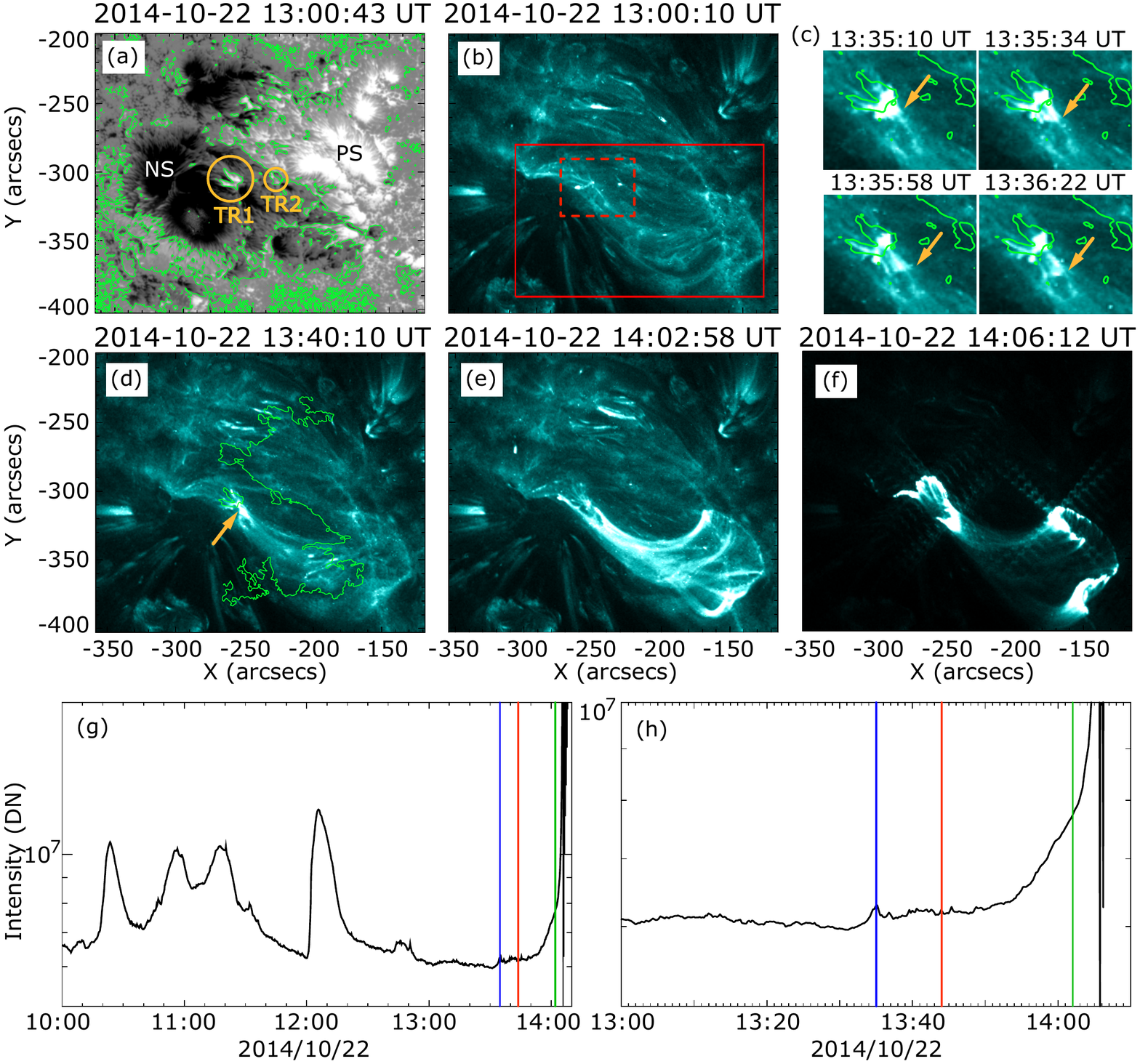}
\caption{
(a) HMI LOS magnetogram with the same FOV of the AIA 131 {\AA} images in (b, d-f).
The format is completely same as the magnetograms in Figure~\ref{fig:HMI}, except that the contours of the strong brightening in AIA images are not over plotted.
(b-f) The time series of the AIA 131 {\AA} images.
The green lines in panels (c, d) represent the PILs such as seen in panel (a).
The small images in panel (c) are enlarged view of the region where is surrounded the red broken line in panel (b).
The yellow arrows indicate the jet-like small eruption and the last precursor brightening in panels (c, d), respectively.
(g, h) The light curve of the AIA 131 {\AA} images.
The total intensity ($DN$ unit) in the region where is surrounded by the red rectangle in panel (b) is plotted.
The vertical green line indicates the X1.6 flare onset time.
The blue and red vertical lines indicate the time when the blueshift and redshift were seen in IRIS \ion{Si}{4} line (Figure~\ref{fig:velocity}), respectively.
%The blue, red, and green vertical lines indicate the time of the blueshift, redshift (seen in Figures~\ref{fig:velocity}, \ref{fig:EIS}), and the X1.6 flare onset, respectively.
Panel (g) covers the time period of 10:00 - 14:10 UT on 2014 October 22, while panel (h) focused on the last one hour before the X1.6 flare onset.
}
\label{fig:AIA131}
\end{figure*}

\begin{figure*}
\epsscale{2.00}
\plotone{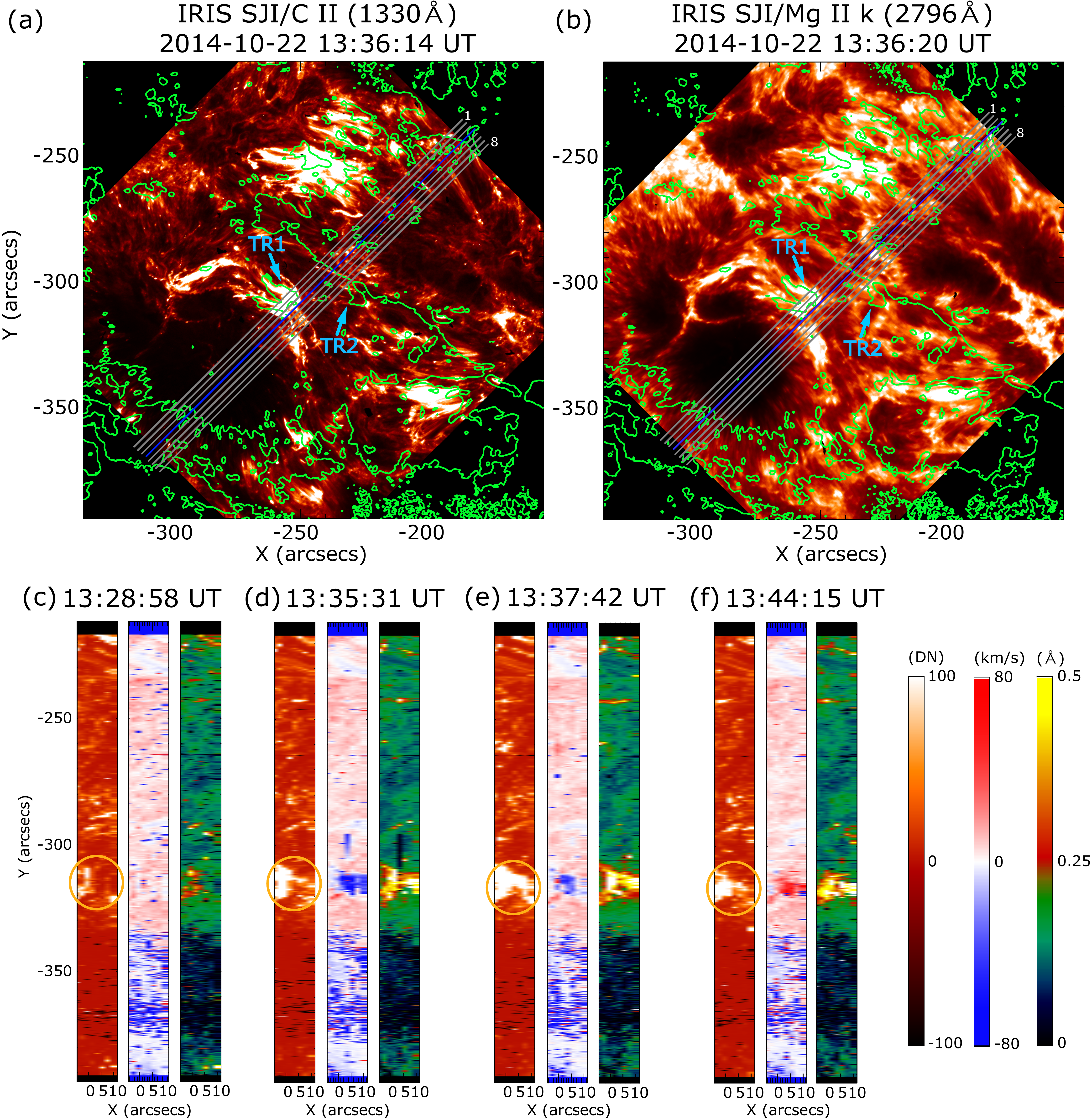}
\caption{
(a, b) IRIS/SJIs of \ion{C}{2} and \ion{Mg}{2} k lines.
Green lines indicate the PILs in HMI LOS magnetogram, and gray lines show the slit positions (only the fourth slit position is represented by the blue line).
The intensity scales are $0$ - $250~DN$ for \ion{C}{2} and $0$ - $800~DN$ for \ion{Mg}{2} k, and TR1 and TR2 are indicated by the sky blue arrows.
(c-f) Intensities (left), Doppler velocities (middle), and line width (right) for \ion{Si}{4} line over TR1 around the time of the last  precursor brightening.
The brightening in TR1, such as seen in Figure~\ref{fig:HMI} is marked by the yellow circle in each intensity image.
(c) No significant Doppler shift is seen over TR1.
(d, e) Blueshift is observed nearby TR1 with the last precursor brightening.
(f) Weak redshift is observed in the same region just after blueshift is observed.
}
\label{fig:velocity}
\end{figure*}

\begin{figure*}
\epsscale{1.50}
\plotone{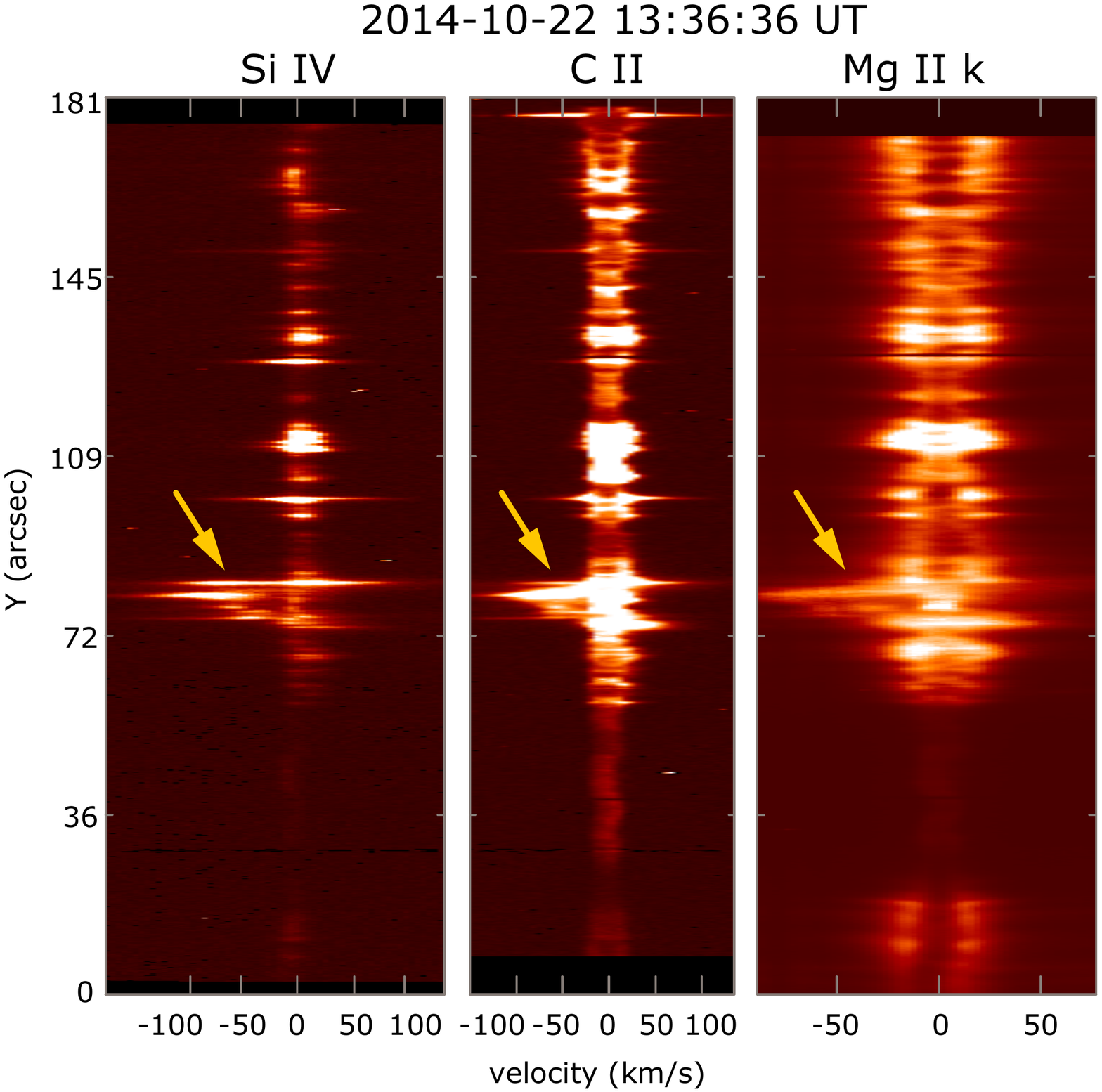}
\caption{
IRIS spectral images of \ion{Si}{4} (left), \ion{C}{2} (middle), and \ion{Mg}{2} k (right) lines.
The spectra were taken at 13:36:36 UT in the fourth slit position which is just over the region where the strong brightening was seen and the magnetic shear was locally opposite to the global magnetic shear along the flaring PIL as seen in Figure~\ref{fig:HMI}.
The reference wavelength for each line $\lambda_{0, Si IV}$, $\lambda_{0, C II}$, and $\lambda_{0, Mg II k}$ are defined as $0~km~s^{-1}$.
Blueshift parts are indicated by the yellow arrows, and the velocities likely more than $-50~km~s^{-1}$ at that time.
}
\label{fig:spectra}
\end{figure*}

The magnetic field in TR1 has characteristic structure.
The magnetic shear locally reverses to the global shear along the flaring PIL as seen in Figure~\ref{fig:HMI}(d).
According to \citet{kusano12}, a small bipole field can be a trigger of a flare when the magnetic shear in the bipole field reverses to the global magnetic shear.
It is so-called a Reversed Shear (RS) type flare-trigger field, and they characterized the condition of the RS-type by two parameters: the shear angle $\theta_{0}$ along the flaring PIL and the azimuth $\varphi_{e}$ of the small bipole field.
In this study, we defined the relative shear angle $\chi$ as the angle between the potential field vector ${\bm B_{p}}$ and horizontal field vector ${\bm B_{h}}$.
The values of $\chi$ along the flaring PIL between PS-NS and $\chi$ in the southwest region of TR1 correspond to the shear angle $\theta_{0}$ and the azimuth $\varphi_{e}$, respectively.
Therefore, our results suggest that the TR1 is the RS-type flare-trigger field of the X1.6 flare.

\citet{kusano12} proposed that an RS-type field can trigger a flare through the local cancellation of magnetic shear of the AR through reconnection with an RS-type of field.
\citet{bamba13} have identified the flare-trigger fields including an RS-type structure for several flare events.
They used precursor brightenings in the chromosphere as a marker of the flare-trigger field.
Their analyses are based on the idea that the chromospheric brightening is caused by preceding magnetic reconnection in the lower atmosphere between a trigger-field and pre-existing sheared field.
They call it ``internal reconnection'' in contrast with ``flare reconnection'' that occurs between the pre-existing sheared fields and that is the main source of the large energy release during a flare.
However, it has been still unclear whether the ``internal reconnection'' is actually occurred in the lower atmosphere such as in the chromosphere.

In this study, we found that all the lines observed by IRIS and Hinode/EIS shows the intensity enhancements, the strong blueshifts (upflows) up to $100~km~s^{-1}$, and the significant line broadenings corresponding to the last precursor brightening over the RS-type magnetic field structure.
The lines are sensitive to emission from middle chromosphere to the corona (log (T) $\sim 4.0$ - $6.3$).
All of these lines show the strong blueshifts up to $\sim100~km~s^{-1}$.
The temperature dependence of these blueshifts is weak.
Generally, pressure gradient driven flows, such as chromospheric evaporation (see Figure~\ref{fig:EIS}(j-l)), show the strong temperature dependence (e.g. \citet{fisher85, watanabe10, imada15}).
Conversely, Lorentz force driven flows, such as reconnection flow, may not show clear temperature dependence. 
Thus, these results suggest that there is an energy input such as magnetic reconnection occurred in lower temperature region than middle chromosphere in which \ion{Mg}{2} k line is sensitive, and that the jet launching chromospheric plasma is observed as the blueshift signals around TR1.
%Note that here we assumed a simple stratified atmosphere and energy input only by thermal conduction.
Magnetic field structure also supports the interpretation that the internal reconnection occurred in lower chromosphere.
When we assume a half-round loop, the loop top reaches around $2,000~km$ altitude because the size of the magnetic field in TR1 is almost $5^{\prime\prime}$, and the internal reconnection might occur from a few hundred to a few thousand kilometer altitude.

\begin{figure*}
\epsscale{2.00}
\plotone{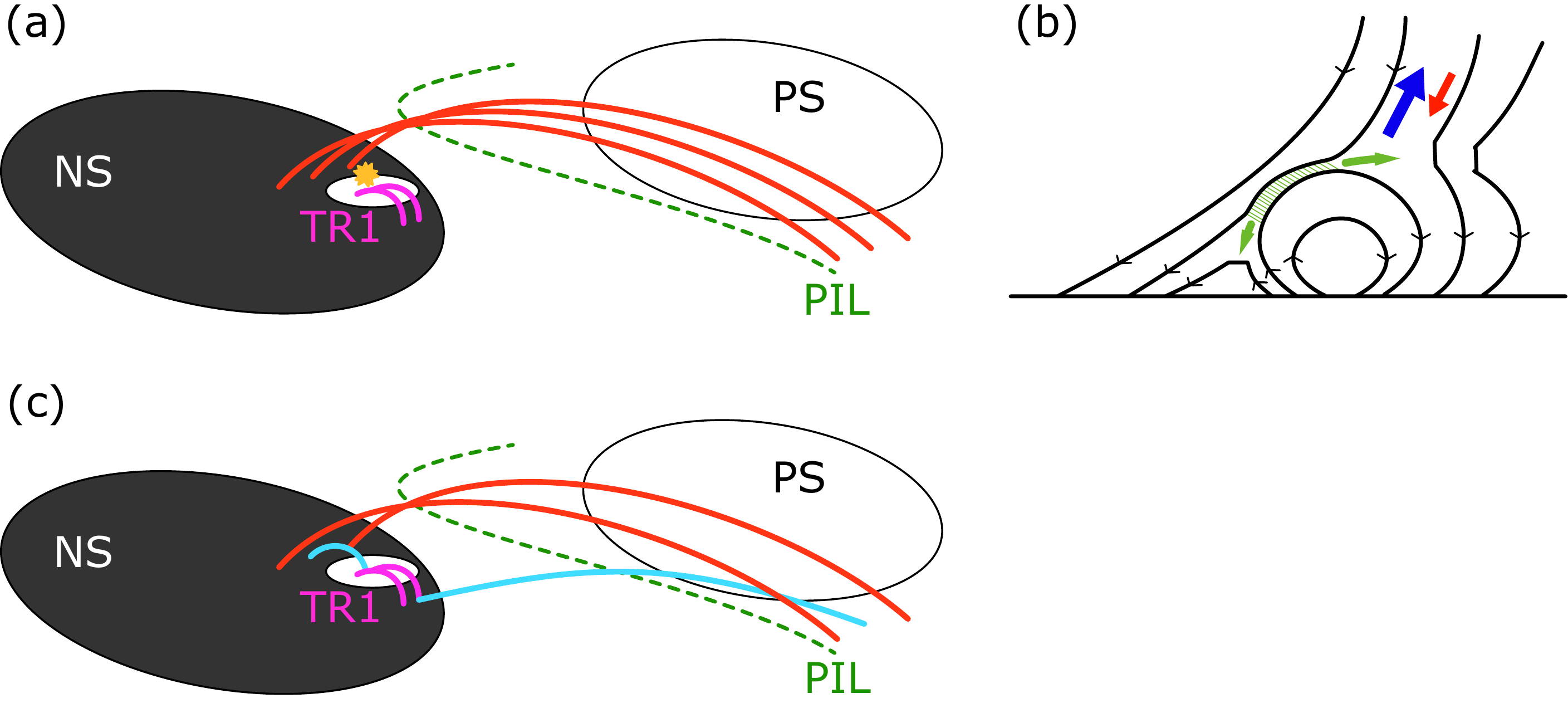}
\caption{
Conceivable scenario of the X1.6 flare in AR 12192.
(a) Shear cancellation (internal reconnection) occurred between TR1 and overlying sheared magnetic field.
(b) Picture of internal reconnection in TR1. Internal reconnection (shear cancellation) occurred in green diagonal line part. Green arrows indicate reconnection jet (comparable to the Alfv$\acute{e}$n velocity). The observed plasma up flow (blueshift) and down flow (redshift) are illustrated by the blue and red arrows, respectively.
(c) Overlying arcades collapse to the region where the magnetic pressure decreased by the shear cancellation, and flare reconnection occurred.
}
\label{fig:scenario}
\end{figure*}

Eventually, we propose the following flare trigger scenario of the X1.6 flare as a picture that is illustrated in Figure~\ref{fig:scenario}.
First, internal reconnection occurred between the overlying sheared magnetic loops (red loops) and small magnetic loops (magenta loops) in TR1, where satisfies the geometrical condition of the RS-type flare trigger region (see panel (a)).
Precursor brightening and blueshift in the chromosphere are observed at that time (approximately $30~min.$ before the flare onset) as seen in Figure~\ref{fig:HMI}(a).
Figure~\ref{fig:scenario}(b) shows the picture of the internal reconnection over TR1, that has similar three-dimensional geometrical condition to the picture of \citet{yokoyama_shibata_95}.%, yokoyama_shibata_96}.
Our results suggest that the internal reconnection occurs at the foot point of the overlying sheared loops (red loops in panels (a)) in the negative polarity region.
Reconnection jets (green arrows), which have the Alfv$\acute{e}$n velocity, may be created both side of the reconnection region (green diagonal line part).
Then the chromospheric plasma is launched upwards as illustrated by the blue arrow, and we observed it as the blueshift with IRIS (Figures~\ref{fig:velocity}(d) and \ref{fig:lineprofiles}(a)) and Hinode/EIS (Figures~\ref{fig:EIS}(d-f) and \ref{fig:EIS_profile}).
After that, the cool materials go down as illustrated by the red arrow, but it might not intrude into the dense and lower chromosphere (formation temperature region of \ion{Mg}{2} k line) because the speed is slower than that of blueshift.
Therefore, \ion{Mg}{2} k line does not significantly redshifted in Figure~\ref{fig:lineprofiles}(b).
The internal reconnection between the red and blue loops cancels magnetic shear over TR1, and it makes small magnetic loops as illustrated by the sky blue lines in panel (c).
Then the overlying sheared loops (red loops) collapse inward the region where the magnetic pressure is decreased by shear cancellation.
Finally, flare reconnection occurs between the red loops and the dense and heated chromospheric plasma evaporates and fills the connecting loop \citep{yokoyama_shibata_01}.
The loop filled by the heated plasma is observed as bright loops in AIA 131 {\AA} (Figure~\ref{fig:AIA131}(e)).
Then the initial flare ribbons appears at the four foot points of the red loops and it enhances as Figure~\ref{fig:HMI}(b, c).
Our scenario is consistent with a possibility discussed in \citet{wallace10}, that reconnection between a newly emerged flux and pre-existing magnetic loops before the flare drives the flows by a pressure gradient between the newly reconnected loops.

Note that the contribution of the other RS-type region TR2 is still unclear.
In Figure~\ref{fig:HMI}(a), we also can see a small brightening in the southwest region of TR2, where $\chi$ is around $90^{\circ}$ (red-colored) in Figure~\ref{fig:HMI}(d).
However, the region is very small in TR2 while that in TR1 is more clearly seen.
Moreover, TR2 locates under the highly sheared magnetic loops along the flaring PIL between PS-NS, whereas TR1 is seemed to locates at almost NS-side foot point of the magnetic loop.
In \citet{kusano12}, they inject an RS-type bipole to just above flaring PIL, then TR2 is more consistent location with their simulation setup.
\citet{bamba_inpress} suggests that an RS-type field could work as a trigger of a flare even if it locates slightly away from the PIL as long as it locates under the highly sheared magnetic loops.
Therefore, it is difficult to identify which TR1 or TR2 is the trigger or both could be a trigger of the X1.6 flare.
In this paper, we have treated only the contribution of TR1 because both IRIS and Hinode/EIS scanned over TR1.

%%%SUMMARY%%%
\section{Summary} \label{sec:summary}

In this study, we studied about the precursor activity for the X1.6 flare on 2014 October 22 in AR NOAA 12192.
We analyzed the photospheric magnetic field structure, the chromospheric/coronal brightening, and the coronal loops from SDO/AIA and HMI data.
We also investigated the spectroscopic data of IRIS and Hinode/EIS that are sensitive to emission from lower chromosphere to the corona.
As a result, we found that the strong blueshift signals corresponding to the last precursor brightening.
The redshifts were also observed just after the blueshifts were observed.
It suggests that the chromospheric plasmas were launched as a jet by magnetic reconnection in the lower chromosphere and fell down.
The photospheric magnetic field structure also suggests that the chromospheric jet was likely caused in the triggering process of the X1.6 flare.
The magnetic shear in the southwest region, where the precursor brightening and chromospheric jet were observed, locally towards the opposite direction to the global magnetic shear along the flaring PIL in the AR.
This feature is consistent with a small bipole field so-called the RS-type field in \citet{kusano12} that can trigger a flare via magnetic shear cancellation with the highly sheared field in the AR.

In this study, we found a secondary flow which seems to be related to reconnection outflow rather than chromospheric evaporation flow. This is not a direct evidence of internal magnetic reconnection between the RS-type field and sheared field of the AR.
%We have not observe bi-directional reconnection outflow that can be a direct evidence of internal magnetic reconnection between the RS-type field and sheared field of the AR.
Nevertheless this study is novelty in the sense that it three-dimensionally investigated the structural change from the photosphere to the corona in the precursor phase, by using both imaging and spectroscopic data. 
Our results show that small photospheric magnetic field structure affects large-scale topological change in the corona via local magnetic reconnection in the chromosphere.
It can support the arguments of \citet{kusano12} that a small magnetic disturbance appears in the photosphere can trigger solar eruptive phenomena such as a flare.
\citet{bamba13} tried to examine the theoretical model by \citet{kusano12} using observational data, and they found small flare triggering field including an RS-type field.
They treated precursor brightening seen in chromospheric line such as \ion{Ca}{2} H as a marker for identification of a flare-trigger field, based on the viewpoint that the precursor brightening caused by magnetic reconnection between a flare-trigger field and sheared field in the AR.
In spite of that, the physical connection between the photosphere to the corona in a flare trigger process was unclear.
Our findings in this study also supports the analysis method to identify the flare trigger region in \citet{bamba13}.
Moreover, in this study, we showed the usefulness of combination observation of imaging and spectroscopy of solar atmosphere and photospheric magnetic field observations.
It is expected that observational and theoretical studies give and take feedbacks each other and that we reveal the details in the physical mechanism and establish the comprehensive understanding of flare trigger.

%%%ACKNOWLEDGEMENTS%%%
\acknowledgments

We would like to offer our special thanks to Drs. David H. Brooks, Tomoko Kawate, Takenori Joten Okamoto, Mr. Ryuichi Kanoh, and people in ISAS/JAXA and NAOJ for the fruitful discussion and comments.
We would like to thank scientific/engineering team of IRIS, SDO, and Hinode for wonderful data.
The HMI and AIA data have been used courtesy of NASA/SDO and the AIA and HMI science teams.
Hinode is a Japanese mission developed and launched by ISAS/JAXA, which collaborates with NAOJ as a domestic partner and with NASA and STFC (UK) as international partners.
Scientific operation of Hinode is conducted by the Hinode science team that is organized at ISAS/JAXA.
This team mainly consists of scientists from institutes in the partner countries.
Support for the post-launch operation is provided by JAXA and NAOJ (Japan), STFC (UK), NASA, ESA, and NSC (Norway).
This work was partly carried out at the NAOJ Hinode Science Center, which is supported by MEXT/JSPS KAKENHI Grant Number 17GS0208, by generous donations from Sun Microsystems, and by NAOJ internal funding.
Part of this work was also carried out on the Solar Data Analysis System (SDAS) operated by the Astronomy Data Center in cooperation with the Hinode Science Center of the NAOJ.
This work was supported by MEXT/JSPS KAKENHI Grant Numbers 	JP16H07478, JP15H05814, JP15H05816, JPG2602, JP23340045, JP26287143, and the joint research program of the Institute for Space-Earth Environmental research (ISEE), Nagoya University.

%%%APPENDIX%%%
%\appendix

%%%References%%%

%\end{document}
%%%TABLES%%%

\begin{table*}
\begin{center}
\begin{tabular}{|c||c|c|c|}
\tableline
Instrument & AIA\tablenotemark{a} & HMI\tablenotemark{b} & HMI\tablenotemark{c} \\
\tableline
Data Type & filtergrams & LOS magnetograms & vector magnetograms \\
\tableline \tableline
FOV & $2000^{\prime\prime} \times 2000^{\prime\prime}$ & $2000^{\prime\prime} \times 2000^{\prime\prime}$ &  $300^{\prime\prime} \times 200^{\prime\prime}$ \\ \hline
Pixel Size & 0.6$^{\prime\prime}$ & 0.5$^{\prime\prime}$ & 0.5$^{\prime\prime}$ \\ \hline
Cadence & 24 sec., 12 sec. & 45 sec. & 12 min. \\ \hline
Wavelength & 1600 {\AA}, 131 {\AA} & 6173 {\AA} & 6173{\AA} \\ \hline
Primary Ion(s) & \ion{C}{4} + continuum, \ion{Fe}{8} + \ion{Fe}{21} & \ion{Fe}{1} & \ion{Fe}{1} \\ \hline
\end{tabular}
\caption{Summary of observation parameters for SDO/AIA and HMI.}
\tablenotetext{a}{{\tt aia.lev1\_uv\_24s} and {\tt aia.lev1\_euv\_12s} series}
\tablenotetext{b}{{\tt hmi.M\_45s} series}
\tablenotetext{c}{{\tt hmi.sharp\_cea\_720s}: SHARP (Spaceweather HMI Active Region Patch) data series}
\label{table:SDO}
\end{center}
\end{table*}
%%%%
\begin{table*}
\begin{center}
\begin{tabular}{|c||c|c|c|}
\hline
Instrument & \multicolumn{2}{c|}{IRIS} & Hinode/EIS \\ \cline{1-4}
Data Type & Slit Jaw Images (SJIs) & Raster Scan & Raster Scan \\ \hline \hline
FOV & 167$^{\prime\prime}$ $\times$ 174$^{\prime\prime}$ & 14$^{\prime\prime}$ $\times$ 174$^{\prime\prime}$ & 59$^{\prime\prime}$ $\times$ 152$^{\prime\prime}$ \\ \hline
Pixel Resolution & \multicolumn{2}{c|}{0.33$^{\prime\prime}$ (FUV), 0.4$^{\prime\prime}$ (NUV)} & 2$^{\prime\prime}$ \\ \hline
Exposure Duration & 15.0 sec. & 14.9 sec. & 9.0 sec. \\ \hline
Cadence (for each step)\tablenotemark{a} & 32.7 sec. & 16.4 sec. & 10.6 Sec. \\ \hline
Wavelength & 1330 {\AA}, 2796 {\AA} & 1330 {\AA}, 1400 {\AA}, 2796 {\AA} & 195.12 {\AA}, 284.18 {\AA}, 256.34 {\AA} \\ \hline
Primary Ion(s) & \ion{C}{2}, \ion{Mg}{2} k & \ion{C}{2}, \ion{Si}{4}, \ion{Mg}{2} k & \ion{Fe}{12}, \ion{Fe}{15}, \ion{He}{2} \\ \hline
Temperature (log (T)) & 4.3, 4.0 & 4.3, 4.8, 4.0 & 6.2, 6.3, 4.9 \\ \hline
Steps & - & 8 positions at 2$^{\prime\prime}$ intervals & 20 positions at 1$^{\prime\prime}$ intervals \\ \hline
Spectral Resolution & - & 50 m{\AA} (FUV), 80 m{\AA} (NUV) & 22 m{\AA} \\ \hline
\end{tabular}
\end{center}
\caption{Summary of observation parameters for IRIS and Hinode/EIS.}
\tablenotetext{a}{The cadence is defined as the difference of the exposure start time between two slit position.}
\label{table:IRIS_EIS}
\end{table*}

%%%FIGURES%%%

%%%%%%%
%%%%%%%
%%%%%%%
%%%%%%%
%%%%%%%
%%%%%%%
%%%%%%%
%%%%%%%
%%%%%%%

\end{document}